\documentclass[journal]{IEEEtran}
\usepackage{amsmath,amsthm,amsfonts,yhmath}
\usepackage{amssymb}
\usepackage[normalem]{ulem}

\usepackage{tikz}
\usepackage{url}
\usepackage{cuted}
\usepackage[nolist]{acronym}
\usepackage{todonotes}

\usetikzlibrary{arrows,shapes}

\usepackage[colorlinks]{hyperref}
\hypersetup{
colorlinks,
    linkcolor={red!40!black},
    citecolor={blue!40!black},
    urlcolor={blue!20!black}
}

\graphicspath{{../figures/}}

\newif\ifgreek
\def\testgreek#1{
  \ifx#1\alpha\greektrue\else
  \ifx#1\beta\greektrue\else
  \ifx#1\gamma\greektrue\else\ifx#1\Gamma\greektrue\else
  \ifx#1\delta\greektrue\else\ifx#1\Delta\greektrue\else
  \ifx#1\epsilon\greektrue\else
  \ifx#1\zeta\greektrue\else
  \ifx#1\eta\greektrue\else
  \ifx#1\theta\greektrue\else\ifx#1\Theta\greektrue\else
  \ifx#1\iota\greektrue\else
  \ifx#1\kappa\greektrue\else
  \ifx#1\lambda\greektrue\else\ifx#1\Lambda\greektrue\else
  \ifx#1\mu\greektrue\else
  \ifx#1\nu\greektrue\else
  \ifx#1\xi\greektrue\else\ifx#1\Xi\greektrue\else
  \ifx#1\pi\greektrue\else\ifx#1\Pi\greektrue\else
  \ifx#1\rho\greektrue\else
  \ifx#1\sigma\greektrue\else\ifx#1\Sigma\greektrue\else
  \ifx#1\tau\greektrue\else
  \ifx#1\upsilon\greektrue\else\ifx#1\Upsilon\greektrue\else
  \ifx#1\phi\greektrue\else\ifx#1\Phi\greektrue\else
  \ifx#1\chi\greektrue\else
  \ifx#1\psi\greektrue\else\ifx#1\Psi\greektrue\else
  \ifx#1\omega\greektrue\else\ifx#1\Omega\greektrue\else
  \ifx#1\varepsilon\greektrue\else
  \ifx#1\vartheta\greektrue\else
  \ifx#1\varrho\greektrue\else
  \ifx#1\varsigma\greektrue\else
  \ifx#1\varphi\greektrue\else
     \greekfalse
  \fi\fi\fi\fi\fi\fi\fi\fi\fi\fi
  \fi\fi\fi\fi\fi\fi\fi\fi\fi\fi
  \fi\fi\fi\fi\fi\fi\fi\fi\fi\fi
  \fi\fi\fi\fi\fi\fi\fi\fi\fi}

\newcommand{\mat}[1]{{\testgreek#1\ifgreek\boldsymbol#1\else
                      \mathbf#1\fi}} % Upright bold for matrices
\newcommand{\ie}{\textit{i.e.}\/, }  % id est ‎(it is).
\newcommand{\eg}{\textit{e.g.}\/, }
\newcommand{\cf}{\textit{cf.}\/, }

\providecommand*{\mrm}[1]{\mathrm{#1}}

\DeclareMathAccent{\ring}{\mathalpha}{operators}{"17}

\providecommand*{\unit}[1]{\ensuremath{\mrm{\,#1}}}  % use eg as $1\unit{m}$
\providecommand*{\ju}{\ensuremath{\mrm{j}}}  % sqrt(-1)

\renewcommand{\Re}{\operatorname{Re}}	% The LaTeX standard is not ISO!
\newlength{\temp}
\newlength{\tempa}

\newcommand{\Um}{\mat{U}}

\newcommand{\Xm}{\mat{X}}

\newcommand{\Jm}{\mat{I}}
\newcommand{\Tm}{\mat{T}}

\newcommand{\Jmt}{\tilde{\mat{I}}}
\newcommand{\Ymt}{\widetilde{\mat{Y}}}

\newcommand{\Rmt}{\widetilde{\mat{R}}}
\newcommand{\Pm}{\mat{P}}

\newcommand{\Hm}{\mat{H}}
\newcommand{\Om}{\mat{0}}

\newcommand{\Zm}{\mat{Z}}
\newcommand{\Mm}{\mat{M}}
\newcommand{\Rm}{\mat{R}}

\newcommand{\Xmm}{\mat{X}_{\mrm{m}}}
\newcommand{\Xme}{\mat{X}_{\mrm{e}}}
\newcommand{\Xmmt}{\widetilde{\mat{X}}_{\mrm{m}}}
\newcommand{\Xmet}{\widetilde{\mat{X}}_{\mrm{e}}}

\newcommand{\Rml}{\mat{R}_{\mrm{\Omega}}}

\newcommand{\xm}{\mat{x}}
\newcommand{\ym}{\mat{y}}
\newcommand{\nm}{\mat{n}}
\newcommand{\MmT}{\wideparen{\Mm}}
\newcommand{\XmeT}{\wideparen{\Xm}_{\mrm{e}}}
\newcommand{\XmmT}{\wideparen{\Xm}_{\mrm{m}}}
\newcommand{\XmT}{\wideparen{\Xm}}
\newcommand{\RmT}{\wideparen{\Rm}}
\newcommand{\RmlT}{\wideparen{\Rm}_{\mrm{\Omega}}}

\newcommand{\Pd}{P_{\mrm{d}}}
\newcommand{\Pl}{P_{\mrm{\Omega}}}

\newcommand{\tran}{\text{T}}
\newcommand{\herm}{\text{H}}

\newcommand{\medel}[1]{\mathcal{E}\left\{ #1\right\}}

\newcommand{\Tr}{\mathop{\mrm{Tr}}\nolimits}

\newcommand{\We}{W_{\mrm{e}}}
\newcommand{\Wm}{W_{\mrm{m}}}

\newcommand{\minimize}{\mrm{minimize}}
\newcommand{\maximize}{\mrm{maximize}}

\newcommand{\subto}{\mrm{subject\ to}}

\newcommand{\Id}{\mat{1}}

\newcommand{\reg}{\varOmega}

\newacro{MoM}[MoM]{method of moments}
\newacro{PEC}[PEC]{perfect electric conductor}
\newacro{EFIE}[EFIE]{electric field integral equation}
\newacro{MFIE}[MFIE]{magnetic field integral equation}
\newacro{FBW}[FBW]{fractional bandwidth}
\newacro{MIMO}[MIMO]{multiple input multiple output}
\newacro{SNR}[SNR]{signal-to-noise ratio}

\title{Fundamental bounds on MIMO antennas}

\author{Casimir Ehrenborg,~\IEEEmembership{Student member,~IEEE}, and Mats Gustafsson,~\IEEEmembership{Member,~IEEE}
\thanks{Casimir Ehrenborg and Mats Gustafsson are with the Department of Electrical and Information Technology, Lund University, Box 118, SE-221 00 Lund, Sweden. (Email: \{casimir.ehrenborg,mats.gustafsson@eit.lth.se\}@eit.lth.se).}% <-this % stops a space
}

\begin{document}

\maketitle

\begin{abstract}
Antenna current optimization is often used to analyze the optimal performance of antennas. Antenna performance can be quantified in \eg minimum Q-factor and efficiency. The performance of MIMO antennas is more involved and, in general, a single parameter is not sufficient to quantify it. Here, the capacity of an idealized channel is used as the main performance quantity. An optimization problem in the current distribution for optimal capacity, measured in spectral efficiency, given a fixed Q-factor and efficiency is formulated as a semi-definite optimization problem. A model order reduction based on characteristic and energy modes is employed to improve the computational efficiency. The performance bound is illustrated by solving the optimization problem numerically for rectangular plates and spherical shells.  
\end{abstract}


\section{Introduction}\label{sec:intro}
Wireless communication in modern systems utilize \ac{MIMO} networks and antennas~\cite{Molisch2011,Paulraj+etal2003}. These systems consist of two sets of antennas, one transmitting, and one receiving. Normally, one of these sets is situated in a location where space allocation is not an issue, such as a base station. However, the other set is usually contained within a small device, such as a mobile phone, where design space is limited~\cite{Ying2012}. Naturally, antenna designs aim at maximizing performance in such an environment. However, there is little knowledge of how the performance depends on size, Q-factor and efficiency restrictions. Having this knowledge \textit{a priori} would enable designers to optimize their antenna designs more efficiently. 
%It is possible to bound performance of small antennas by creating optimal current distributions given a set of design restrictions. However, this has, so far, not been successfully calculated for generic structures. 
There has been efforts to bound \ac{MIMO} antennas performance for spherical surfaces~\cite{AlayonGlazunov+etal2011,Gustafsson+Nordebo2007a} and through information-theoretical approaches~\cite{Migliore2008,Taluja+Hughes2012,Kundu2016}. In this letter a method for constructing a performance bound on capacity for arbitrary shaped \ac{MIMO} antennas using current optimization is presented.

Antenna current optimization can be used to determine physical bounds for antennas of arbitrary shape~\cite{Gustafsson+etal2015b}. These physical bounds are found by maximizing a certain performance parameter by freely placing currents in the design space. By having total control of the current distribution an optimal solution can be reached. While these currents might not necessarily be realizable they provide an upper bound for the considered problem. Construction of such physical bounds are made possible by the ability to formulate convex optimization problems~\cite{Boyd+Vandenberghe2004} for the performance quantity of interest. The performance of simple antennas can be quantified in \eg the Q-factor, gain, directivity, and efficiency~\cite{Gustafsson+etal2016a}. \ac{MIMO} antennas, on the other hand, are more complex and a single parameter is insufficient to determine their performance. As such, it is a challenging problem to construct physical bounds for \ac{MIMO} systems. However, it is still possible to utilize antenna current optimization to maximize a given performance quantity, such as capacity, with restrictions on, \eg the Q-factor and efficiency.

In communication theory a \ac{MIMO} network's capacity is usually optimized for a fixed set of antennas. The performance of the antennas is accepted as it is and the upper bound on network performance is calculated by \eg water filling~\cite{Paulraj+etal2003}. However, in doing so we forgo an opportunity to gain extra performance through optimizing the antennas. In this paper we illustrate how bounds on capacity of a \ac{MIMO} antenna can be determined by antenna current optimization.

Considering the channel between two sets of antennas leads to optimization for specific scenarios or circumstances, in this paper we are interested in establishing general performance bounds for \ac{MIMO} antennas. As such we focus on one set of antennas and idealize the other. The second set of antennas are characterized as the spherical modes in the far-field. This leads to an idealized channel in terms of spherical modes~\cite{Gustafsson+Nordebo2007a}, which can be thought of as a direct line of sight channel where all radiation is received. Considering such a channel also has the benefit of reducing computational complexity. This is further reduced by a model order reduction of the \ac{MoM} impedance matrix characterizing the antenna.

The convex optimization problem is constrained by the efficiency or Q-factor. These are expressed as quadratic forms in the current density, where the stored energy in \cite{Vandenbosch2010} is used. This leads to a convex optimization problem that maximizes the capacity in terms of spectral efficiency for a fixed \ac{SNR} and Q-factor. The convex optimization problem is a semi-definite program~\cite{Boyd+Vandenberghe2004}  expressed in the covariance matrix of the current distribution.

\section{MIMO model} \label{sec:MIMO}
\begin{figure}%
\centering
\includegraphics[width=\columnwidth]{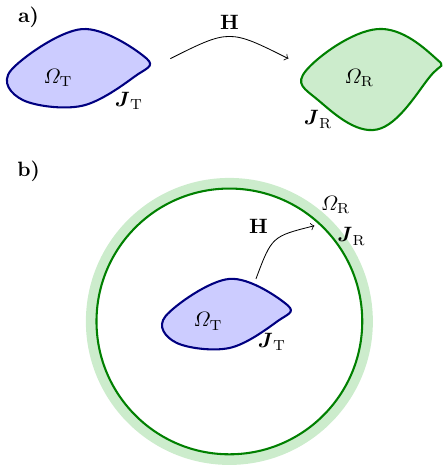}
\caption{Illustration of the \ac{MIMO} system model with transmitter region $\reg_{\mrm{T}}$ and receiver region $\reg_{\mrm{R}}$. Part \textbf{(a)} shows the classical \ac{MIMO}  setup with spatially separated regions. Part \textbf{(b)} illustrates the idealized case when the receiver region entirely surrounds the transmitter. The system in \textbf{(b)} is utilized in this paper to determined performance bounds on \ac{MIMO} antennas confined to the region $\reg_{\mrm{T}}$.}%
\label{fig:MIMOsyst}%
\end{figure}

A classical \ac{MIMO} system is modeled as~\cite{Paulraj+etal2003}
\begin{equation}\label{eq:MIMOmodel}
	\ym = \Hm\xm + \nm ,
\end{equation}
where $\xm$ is a $N\times 1$ matrix of the input signals, $\ym$ is a $M\times 1$ matrix of the output signals, $\nm$ is a $M\times 1$ matrix of additive noise, and $\Hm$ is the $M\times N$ channel matrix. The channel matrix models how power is transmitted from the input signals to the output signals, this includes the receiving and transmitting antennas and the wave propagation between them~\cite{Paulraj+etal2003}. 

Fig.~\ref{fig:MIMOsyst}a displays a classical \ac{MIMO} setup where two sets of antennas form a channel. Analysis of such systems depend greatly on external factors, such as, scattering phenomena, channel characterization, and antenna location~\cite{Paulraj+etal2003}. However, to investigate performance bounds for \ac{MIMO} antennas we must limit the degrees of freedom to a single antenna. This implies that $\Hm$ in~\eqref{eq:MIMOmodel} should model the channel between an arbitrary antenna and an idealized receiver, corresponding to Fig.~\ref{fig:MIMOsyst}b. 
The transmitting antenna is modeled with its current distribution using a \ac{MoM} approximation~\cite{Gustafsson+etal2016a} such that each basis function corresponds to an element of $\xm$. The receiver is modeled with the radiated spherical modes, where each mode is an element in $\ym$~\cite{Gustafsson+Nordebo2006c,Gustafsson+Nordebo2007a}. This leads to a \ac{MIMO} system of infinite dimension as $N$ increases with mesh refinement and $M$ increases with the number of included spherical modes. In numerical evaluation $N$ and $M$ are chosen sufficiently large to ensure convergence.  

The transmitted signals are modeled as the \ac{MoM} current elements $\Jm=\Tm\xm$, where the matrix $\Tm$ maps the transmitted signals $\xm$ to the current distribution on the antenna $\Jm$. The covariance matrix of the transmitted signal is $\Pm=\frac{1}{2}\medel{\xm\xm^{\herm}}$,
where $\medel{\cdot}$ denotes the temporal average~\cite{Paulraj+etal2003}. With this matrix we can calculate the average transmitted power,
\begin{multline}
  P = \frac{1}{2}\medel{\Jm^{\herm}\Rm\Jm} 
  = \frac{1}{2}\medel{\xm^{\herm}\Tm^{\herm}\Rm\Tm\xm}\\ 
  = \frac{1}{2}\Tr\medel{\Tm^{\herm}\Rm\Tm\xm\xm^{\herm}}
  = \Tr(\RmT\Pm) ,
\end{multline}  
where $\RmT=\Tm^{\herm}\Rm\Tm$, and $\Rm$ is the resistive part of the \ac{MoM} impedance matrix, $\Zm = \Rm +\ju\Xm$~\cite{Gustafsson+etal2016a}.
Since we are concerned with connecting the currents on the antenna structure to the spherical modes~\cite{Gustafsson+Nordebo2013} in the idealized receiver we express our channel as
\begin{equation}
  \ym  = \Mm\Jm +\nm= \Mm\Tm\xm +\nm
  =\MmT\xm +\nm ,
\label{eq:receiverMode}
\end{equation}
where $\Mm$ denotes the map from the currents to the spherical modes. This is a direct channel between the antenna current distribution and the spherical modes~\cite{Gustafsson+Nordebo2006}. 
The capacity, expressed as spectral efficiency~($\unit{b/(s\,Hz)}$), of this channel is given by~\cite{Paulraj+etal2003}
\begin{equation}\label{eq:RayECap}
  C
  = \max_{\Tr(\RmT\Pm)=P}
  \log_2\det\left(\Id+\frac{1}{N_0}
      \MmT
      \Pm                                        
      \MmT^{\herm}
    \right),
\end{equation}
where $\Id$ is the $M\times M$ identity matrix, and $N_0$ is the noise power. The noise is modeled as white complex Gaussian noise. % with covariance matrix $\medel{\nm\nm^{\herm}}=\Id_{M} N_0$. 
The optimal energy allocation in this channel for capacity maximization is given by the water-filling solution~\cite{Paulraj+etal2003}. Alternatively, the optimal solution for this problem can be solved by a semidefinite optimization program,
\begin{equation}\label{eq:convexGQ_minFJ}
\begin{aligned}
	& \maximize && \log_2\det(\Id+\gamma\MmT\Pm\MmT^{\herm}) &&&\\
	& \subto && \Tr(\RmT\Pm) = 1 &&& \\
  & && \ \Pm \succeq 0 ,&&&
\end{aligned}	
\end{equation}
where the unit transmitted power is considered, and $\gamma=P/N_0$ is the total \ac{SNR}. Maximizing the capacity of this channel corresponds to focusing the radiation of the antenna to the orthogonal spherical modes.

The solution to~\eqref{eq:convexGQ_minFJ} is unbounded and increases as mesh refinement and the number of spherical modes are increased if the \ac{SNR} is scaled with the number of channels in $\MmT$~\cite{Paulraj+etal2003}. Here, we consider the case of a fixed \ac{SNR} where the solution only depends on the \ac{SNR}~\cite{TEAT-7265,TEAT-7266}.
The solution to~\eqref{eq:convexGQ_minFJ} can be made more realistic by adding constraints on the losses or Q-factor of the transmitting antenna~\cite{Gustafsson+Nordebo2007a,Morris+etal2005}. The Ohmic losses are calculated as
\begin{equation}
  \Pl = \frac{1}{2}\medel{\Jm^{\herm}\Rml\Jm}
  =\frac{1}{2}\medel{\xm^{\herm}\Tm^{\herm}\Rml\Tm\xm}
  =\Tr(\RmlT\Pm) , 
\end{equation}
where $\RmlT=\Tm^{-\herm}\Rml\Tm$, and $\Rml$ is the loss matrix of the antenna~\cite{Gustafsson+etal2016a}. The stored electric energy is
\begin{equation}
  \We = \frac{1}{4\omega}\medel{\Jm^{\herm}\Xme\Jm}
  =\frac{1}{4\omega}\medel{\xm^{\herm}\Tm^{\herm}\Xme\Tm\xm}
  %=\frac{1}{4\omega}\Tr\Xmet\medel{\xm\xm^{\herm}}
  =\frac{1}{2\omega}\Tr(\XmeT\Pm) ,
\end{equation}
where $\XmeT=\Tm^{\herm}\Xme\Tm$, and $\Xme$ is the electric reactance matrix~\cite{Gustafsson+etal2016a}. The stored magnetic energy $\Wm$ is similarly defined by the magnetic reactance matrix $\Xmm$ as $\Wm=\frac{1}{2\omega}\Tr(\XmmT\Pm)$, where $\XmmT=\Tm^{\herm}\Xmm\Tm$.

With these constraints in hand we can formulate our optimization problem. We note that the solution is independent of the power $P$, so it is sufficient to consider the case $P=1$ giving
\begin{equation}\label{eq:convex_MIMOr}
\begin{aligned}
	& \maximize && \log_2\det(\Id+\gamma\MmT\Pm\MmT^{\herm})\\
	& \subto &&  \Tr((\XmeT+\XmmT)\Pm) \leq 2Q \\
	& && \Tr(\XmT\Pm) = 0 \\
	& && \Tr(\RmlT\Pm) \leq 1-\eta \\
	& && \Tr(\RmT\Pm) = 1 \\
	& && \ \Pm \succeq \Om,
\end{aligned}	
\end{equation}
where $\eta$ is the antenna efficiency, and self-resonance is enforced. Here, the problem has been normalized to dissipated power, including losses. The consequence of this is that the Q-factor considered includes losses in its calculation. It is possible, and sometimes advantageous, to normalize to different quantities such as the radiated power. Equation~\eqref{eq:convex_MIMOr} is a semi-definite optimization problem which has a unique solution~\cite{Boyd+Vandenberghe2004}. However, the problem is non-trivial due to the large number of unknowns for realistic antenna problems. For example a rectangular plate of size $\ell\times\ell/2$ discretized into $64\times 32$ rectangular elements has $N=4000$ unknowns. This size is not a problem for convex optimization of type $G/Q$ and $Q$~\cite{Gustafsson+Nordebo2013,Capek+etal2017b,Gustafsson+etal2016a}. However, the semi-definite relaxation has close to $N^2/2=8\cdot 10^6$ unknowns, making the problem much more computationally demanding. Moreover, the logarithm used in the definition of capacity is more involved than the simple quadratic functions in $G/Q$ and $Q$ type problems~\cite{Gustafsson+Nordebo2013,Gustafsson+etal2016a}. Here, the number of unknowns is reduced by expansion of the currents in characteristic, energy, and efficiency modes~\cite{Gustafsson+etal2016a}, with similar results. 

The expansion includes only the dominating modes and as such constitutes a model order reduction. This implies a change of basis $\Jm  \approx  \Um\Jmt$,
where $\Um$ maps between the old and the new currents. This reduces the number of unknowns to the included modes $N_1\ll N$. With this approximation the stored energy, for example, is calculated as
\begin{multline}
  \Jm^{\herm}\XmeT\Jm
  \approx\Jmt^{\herm}\Um^{\tran}\XmeT\Um\Jmt
  =\Jmt^{\herm}\Xmet\Jmt \\
  =\Tr(\Xmet\Jmt\Jmt^{\herm})
  =\Tr(\Xmet\Ymt) , 
\end{multline}
where $\Ymt=\Jmt\Jmt^{\herm}$, and $\Xmet=\Um^{\tran}\XmeT\Um$. Similarly $\XmmT$, and $\RmT$, are expressed as $\Xmmt = \Um^{\tran}\XmmT\Um$, and $\Rmt = \Um^{\tran}\RmT\Um$. These replace the corresponding matrices in~\eqref{eq:convex_MIMOr}, with $\Ymt$ replacing $\Pm$. This reduces the number of unknowns from approximately $N^2/2$ to $N_1^2/2$.

\section{Numerical examples}
In the following examples the optimization problem~\eqref{eq:convex_MIMOr} has been solved for a \ac{MIMO} system resembling Fig.~\ref{fig:MIMOsyst}b using the Matlab library CVX~\cite{Gustafsson+etal2016a,Grant+Boyd2011}. The logarithm in the optimization problem~\eqref{eq:convex_MIMOr} was replaced by a root of order $M$~\cite{Grant+Boyd2011}. After the optimization has been carried out the capacity is calculated as normal with the optimized currents. The energy restriction on the number of transmitter modes and the number of spherical harmonic modes in the receiver have been chosen sufficiently large to ensure convergence and varies from example to example. Using too many modes may also result in the solver failing to solve the problem due to its size and must therefore be regulated for each run individually. 
Since the performance of a \ac{MIMO} antenna cannot be quantified by a single parameter the optimization was run with different constraints. This illustrates how capacity is bounded by different requirements on the transmitting antennas.
The optimization has also been run for a spherical shell circumscribing the antenna.

\begin{figure}%
\centering
\includegraphics[width=0.9\columnwidth]{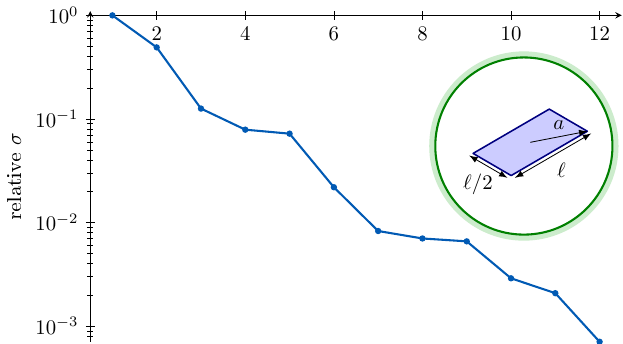}%
\caption{The singular values of the channel matrix $\MmT$ for a rectangular plate $\ell\times\ell/2$ for the wavelength $\ell=0.21\lambda$.}%
\label{fig:SingValplate}%
\end{figure}
\begin{figure}%
\centering
\includegraphics[width=0.9\columnwidth]{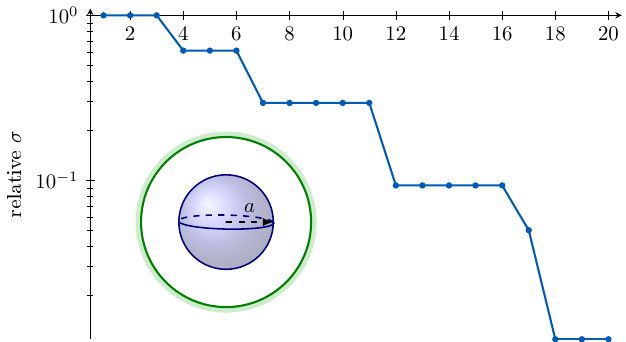}%
\caption{The singular values of the channel matrix $\MmT$ for a spherical shell $r=a$, where $a=0.56\ell$, for the wavelength $\ell=0.21\lambda$.}%
\label{fig:SingValsph}%
\end{figure}
By performing a singular value decomposition of the channel matrix $\MmT$ we can see how many channels dominate the information transfer between the plate and the spherical modes, see Fig.~\ref{fig:SingValplate}. Here, we see that there are only a few channels that dominate the rest. This indicates that so long as our model order reduction preserves these channels it produces correct solutions.

\begin{figure}%
\centering
\includegraphics[width=0.9\columnwidth]{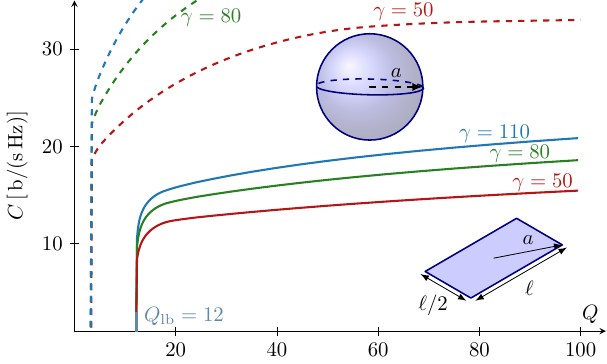}%
\caption{Maximum spectral efficiency achievable for a loss-less rectangular plate of size $\ell\times\ell/2$ for the wavelength $\ell=0.21\lambda$ given maximum Q-factor on the horizontal axis. The dashed lines show the maximum spectral efficiency achievable for the corresponding circumscribing spherical shell.}%
\label{fig:CapvsQ_SNR}%
\end{figure}
In Fig.~\ref{fig:CapvsQ_SNR} the capacity has been optimized for a plate of electrical size $\ell=0.21\lambda$, and is depicted as a function of the Q-factor restriction. We see a cut-off for $Q\leq12$ where the optimization problem is unable to realize a feasible current distribution for so low Q-factor, \cf the lower bound on the Q-factor~\cite{Capek+etal2017b}. For higher \ac{SNR} the capacity increases but the cut-off stays the same, since the \ac{SNR} does not affect the Q-factor.

\begin{figure}%
\centering
\includegraphics[width=0.9\columnwidth]{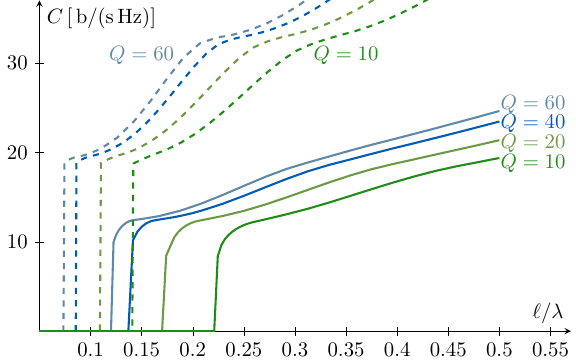}%
\caption{Maximum spectral efficiency achievable for a loss-less rectangular plate of electrical size $\ell/\lambda$ for maximum Q-factor with \ac{SNR} $\gamma=50$, \cf Fig.~\ref{fig:CapvsQ_SNR}. The dashed lines show the maximum spectral efficiency achievable for the corresponding circumscribing spherical shell. }%
\label{fig:Capvs_L_Q}%
\end{figure}
We can instead regard the problem with a fixed \ac{SNR} and investigate how the capacity varies with antenna size, see Fig.~\ref{fig:Capvs_L_Q}. Depending on which $Q$ is chosen the solution is only realizable for sizes above a certain cut-off. This cut-off corresponds to the size which has the chosen $Q$ as its minimum achievable $Q$. Above this size the capacity seems to depend linearly on the antenna size. 
This is consistent with how capacity scales with the number of antennas included in a \ac{MIMO} system~\cite{Paulraj+etal2003}. 

In both Fig.~\ref{fig:CapvsQ_SNR} and~\ref{fig:Capvs_L_Q} the dashed lines show the optimization problem solved for a spherical shell circumscribing the planar region. We see that the spectral efficiency achievable by a planar antenna is much less than that of the sphere.

\begin{figure}%
\centering
\includegraphics[width=0.9\columnwidth]{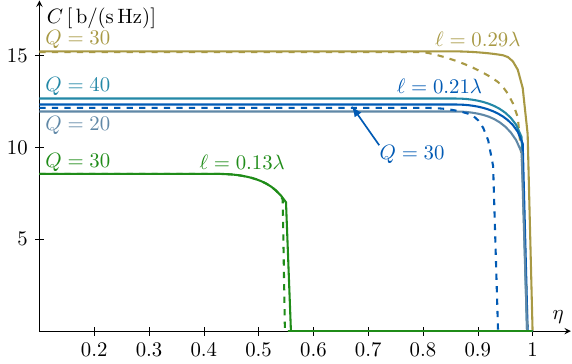}%
\caption{Maximum spectral efficiency achievable for a rectangular plate of electrical size $\ell/\lambda$ for minimum efficiency $\eta$. The losses are modeled as a resistive sheet with $R=0.2\unit{\Omega}/\Box$. The minimum Q-factor is set to 30 for the three main graphs and \ac{SNR} $\gamma=50$. Solid lines are optimized without enforcing resonance and dashed lines are optimized with resonance. For $\ell=0.21\lambda$ the Q-factors $[20,30,40]$ are plotted.}%
\label{fig:CapvsEff_L_Q}%
\end{figure}
Setting an efficiency requirement on the optimization may restrict which modes are realizable. Fig.~\ref{fig:CapvsEff_L_Q} illustrates how capacity varies as a function of antenna efficiency. We see that the capacity is unaffected until some cut-off value where the solution is no longer realizable. For electrical sizes $\ell=0.21\lambda~\text{and}~0.29\lambda$ this occurs when antenna efficiency requirements is high, above $90\%$. However, for smaller sizes, such as $\ell=0.13\lambda$, we see that this cut-off occurs at lower antenna efficiencies. The optimization problem has been solved both with and without enforcing resonance. When resonance is enforced, showed in dashed lines, we see that the cut-off occurs at lower efficiencies, this is due to self-resonant currents being inherently less efficient~\cite{Jelinek+Capek2017}. For the size $\ell=0.21\lambda$ the Q-factor requirement was varied as well, leading to a slight reduction or increase in capacity. Close to the cut-off efficiency we see a slight decrease in capacity for all cases. This corresponds to the requirement on efficiency limiting the optimization problem. For lower efficiency requirements other constraints limit the optimization and the capacity is unaffected by the bound on efficiency.

\begin{figure}%
\centering
\includegraphics[width=1\columnwidth]{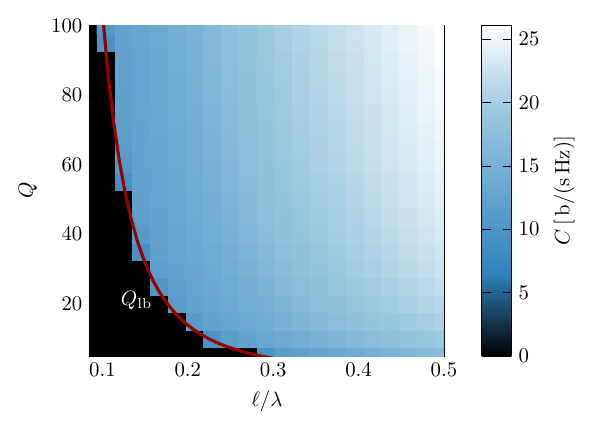}%
\caption{Illustration of the bounding surface of spectral efficiency for a loss less rectangular plate as a function of size and Q-factor with \ac{SNR} $\gamma=50$. The red curve shows minimum $Q$~\cite{Capek+etal2017b}.}%
\label{fig:CapvsL_Q}%
\end{figure}
In Fig.~\ref{fig:CapvsL_Q} both the size of the antenna and the Q-factor are varied to create a two dimensional bounding surface. This surface has a sharp cut-off along the minimum $Q$ line~\cite{Capek+etal2017b} seen on the left in Fig.~\ref{fig:CapvsL_Q}. We see that the increase in capacity follows the shape of the minimum $Q$ curve as $\ell/\lambda$ and $Q$ are increased. 
This surface provides a bound on the capacity achievable for \ac{MIMO} antennas of different sizes and with different bandwidth requirements.

\section{Conclusions}\label{S:conclusions}
In this letter we have presented a framework for constructing performance bounds for \ac{MIMO} antennas. We simplified the channel problem often considered in communication theory to an idealized channel consisting of a spherical receiver surrounding the antenna region.
This enables the formulation of a semi-definite optimization problem that gives a bounding capacity for any antenna that can be constructed within the considered region limited by size, \ac{SNR}, antenna efficiency, and Q-factor. By utilizing a model order reduction based on energy and characteristic modes~\cite{Capek+etal2017b} the complexity of the problem is reduced such that it is solvable.  

These physical boundaries of \ac{MIMO} antennas represent the ideal solutions possible given complete freedom of current placement within the design area. 
While the shape of these current distributions are not easily realizable~\cite{Jelinek+Capek2017}, the bounding values provide an upper limit to what is possible for real antenna topologies.
%While such a current distribution may not be realizable explicitly~\cite{Jelinek+Capek2017}, the bounding values are useful as a tool in antenna design. 
It remains interesting to investigate how these bounds compare to antenna designs and measurements.

\section*{Acknowledgment}
The support of the Swedish foundation for strategic research under the program applied mathematics and the project Complex analysis and convex optimization for electromagnetic design is gratefully acknowledged.

\begin{appendices}
\section{Antenna parameters}\label{App:antennapar}
The impedance matrix $\Zm=\Rm+\ju\Xm$ is determined from a MoM description of the antenna structure. The impedance matrix is divided into its resistance $\Rm$ and reactance $\Xm$. Moreover, the reactance is decomposed into its magnetic and electric parts, \ie $\Zm = \Rm + \ju (\Xmm - \Xme)$, where the stored electric and magnetic energies are~\cite{Vandenbosch2010,Cismasu+Gustafsson2014a}
\begin{subequations}
\label{eq:WDef}
\begin{equation}
	\Wm \approx \frac{1}{8}\Jm^\herm \left( \frac{\partial\Xm}{\partial\omega} + \frac{\Xm}{\omega} \right)\Jm = \frac{1} {4\omega}\Jm^\herm\Xmm\Jm,
\label{eq:WmDef}
\end{equation}
\begin{equation}
	\We \approx \frac{1}{8}\Jm^\herm \left( \frac{\partial\Xm}{\partial\omega} - \frac{\Xm}{\omega} \right)\Jm = \frac{1} {4\omega}\Jm^\herm\Xme\Jm,
\label{eq:WeDef}
\end{equation}
\end{subequations}
respectively, and the dissipated power $\Pd$ is given by
\begin{equation}
	\Pd = \frac{1}{2} \Jm^\herm \Rm \Jm.
\label{eq:PrDef}
\end{equation}
The Q-factor is defined as the quotient between the time-average stored and dissipated energies~\cite{Ohira2016,Volakis+etal2010,Gustafsson+etal2015b}
\begin{equation}
Q = \frac{2 \omega \max\{\We, \Wm \}}{\Pd} 
%= \max \left\{ Q_{\mrm{e}}, Q_{\mrm{m}} \right\} \\
=\frac{\max\{\Jm^\herm\Xme\Jm,\Jm^\herm\Xmm\Jm\}}{\Jm^\herm \Rm \Jm}.
\label{eq:Qdef}
\end{equation}

\section{Maximum efficiency}
To motivate the cut-off values seen in Fig.~\ref{fig:CapvsEff_L_Q} the maximum efficiency for given Q-factors was investigated. This was evaluated using two optimization problems, one to find the minimum efficiency for a set Q,
\begin{equation}\label{eq:convex_EffQ}
\begin{aligned}
	& \minimize && \Re\Tr(\RmlT\Pm)\\
	& \subto &&  \Tr((\XmeT+\XmmT)\Pm) = 2Q \\
	& && \Tr(\XmT\Pm) = 0 \\
	& && \Tr(\RmT\Pm) = 1 \\
	& && \ \Pm \succeq \Om,
\end{aligned}	
\end{equation}
and one to find the minimum Q-factor for a certain efficiency,
\begin{equation}\label{eq:convex_QEff}
\begin{aligned}
	& \minimize && \Re\Tr((\XmeT+\XmmT)\Pm)\\
	& \subto &&  \Tr((\XmmT-\XmeT)\Pm) = 0 \\
	& && \Tr(\RmT\Pm) = 1 \\
	& && \Tr(\RmlT\Pm) = 1-\eta\\  
	& && \ \Pm \succeq \Om.
\end{aligned}	
\end{equation}
These problems can be reformulated so that resonance is not enforced, this results in a higher efficiency limit. The optimization problems~\eqref{eq:convex_EffQ} and~\eqref{eq:convex_QEff} for the efficiency also arise from semi-definite relaxation~\cite{Boyd+Vandenberghe2004} of the corresponding problems formulated in the current $\Jm$. Semi-definite relaxation is a technique to solve quadratically constrained quadratic programs (QCQP) and can applied to many antenna problems~\cite{Luo+etal2010,Fuchs2014,Jonsson+etal2017}.

\end{appendices}

% Generated by IEEEtran.bst, version: 1.14 (2015/08/26)

\end{document}